\documentclass[12pt]{article}
\usepackage{graphicx}
\usepackage{lscape}
\usepackage{eurosym}
\bibstyle{plain}

\newcommand{\be}{\begin{equation}}
\newcommand{\en}{\end{equation}}

\newcommand{\PP}{{\mathord{I\kern -.33em P}}}
\newcommand{\EE}{{\mathord{I\kern -.33em E}}}
\newcommand{\RR}{{\mathord{I\kern -.33em R}}}

\newcommand{\ea}{\end{eqnarray}}
\newcommand{\ba}{\begin{eqnarray}}
\newcommand{\ean}{\end{eqnarray*}}
\newcommand{\ban}{\begin{eqnarray*}}

\begin{document}

\title{A heuristic pricing and hedging framework for multi-currency fixed income desks} 

\author{Eduard Gim\'{e}nez \thanks{Head of Model Development Group, Front Office, CaixaBank, Av. Diagonal, 621-629 T.I. P13, Barcelona, Spain {\em eduard.gimenez.f@lacaixa.es}.} \and Alberto Elices\thanks{Head of XVA Model Validation, Model Risk, Bank Santander, Av. Cantabria s/n, 28660 Boadilla del Monte, Spain, {\em aelices@gruposantander.com}.} \and Giovanna Villani \thanks{Quantitative Analyst, Front Office, CaixaBank, Av. Diagonal, 621-629 T.I. P13, Barcelona, Spain, {\em gvillani@lacaixa.es}.}. }
\date{\today}
\maketitle

\begin{abstract}
{\small 
It is well known that traded foreign exchange forwards and cross currency swaps (CCS) cannot be priced applying overnight cash and carry arguments as they imply absence of funding advantage of one currency to the other. This paper proposes a heuristic present value concept for multi-currency pricing and hedging which allows taking into account the funding and therefore the collateral currency and its pricing impact. For uncollateralized operations, it provides more funding optionality to achieve either cheaper or more connected funding to the hedging instruments. When foreign exchange forwards get aligned with overnight cash and carry arguments, this method naturally converges to the well established OIS discounting where each leg is funded in its own currency. A worked example compares this approach with a benchmark.}
\end{abstract}

\section{Introduction}
\label{sec:Introduction}

Before the financial crisis started in July 2007 with Bear Stearns default, interest rate desks would essentially use a unique interest rate curve for each currency to price and hedge derivative products. The crisis showed that big investment banks could default and therefore lending and borrowing activities became severely restricted no longer allowing for certain hedging strategies. Ever since, basis spreads of tenor swaps where no longer negligible.

After some debate, today there exists a broad consensus about how this issue must be handled for a single currency setting. Regarding different tenors, i.e. 3 and 6 month FRA (Forward Rate Agreement), as independent entities, leaves enough room for new variables to fit the arbitrage free equations that traded products must satisfy. In the single currency setting (see \cite{Bianchetti2009b}), the discount curve for each currency is build out of overnight indexed swaps (OIS) through a bootstrapping algorithm. Forward Libor estimation curves for each different tenor frequency are thereafter calculated out of par and tenor swap quotes by another bootstrapping algorithm based on the previously calculated discount curve. This setting is simply called OIS discounting and will be referred in this way throughout this paper. 

In the cross currency swap market a similar situation exists. The price of traded foreign exchange (FX) forwards and cross currency swaps (CCS) cannot be exactly derived from arbitrage free models calibrated exclusively to swap markets on each currency and the FX spot price,  (see \cite{Cuesta2012} for a heuristic analytic formula to predict basis tenor and cross currency swaps). This is so, because FX forward and CCS prices are driven, among other things, by trading flows and as long as the differences between the ``theoretical'' and market prices remain low, the expected return will not be enough to compensate the bank capital expenses. Unfortunately, for the multi-currency setting, it is not easy to introduce new variables in a similar way as it was for the single currency setting (see \cite{Fujii2010}).

The paper proposes a heuristic present value concept (applied to deterministic cash flows) for multi-currency pricing and hedging illustrated for cross currency swaps. This naturally allows choosing the funding and therefore the collateral currency in the valuation and its pricing impact. Starting from OIS discounting in each currency, the approach is based on the decomposition of customized cross currency swaps as a combination of market quoted cross currency swaps plus a minor amount of additional cash flows which embed the pricing and hedging components which actually are model dependent. These minor additional cash-flows may then be funded in the collateral currency for collateralized operations. In addition, for uncollateralized operations these cash-flows may be funded in the cheapest currency or in the currency which provides the best connection with the products used for hedging. When market foreign exchange forwards are replicated by overnight cash and carry arguments\footnote{Overnight cash and carry arguments imply FX forwards calculated with OIS discount factors in both currencies. This implicitly assumes that funding in both currencies is symmetric and there is not a funding advantage of one currency with respect to the other.}, this heuristic multi-currency pricing converges to OIS cash flow discounting in the currency in which cash flows are denominated. This heuristic framework is valid only for the most frequently traded products (e.g. swaps and cross currency swaps). Valuation of complex multi-currency exotic products such as callable swaps are beyond the scope of this framework.

When the paper assumes full collateralization of trades (see \cite{Bianchetti2009b}), it also considers a funding structure between Front Office desks and the balance sheet of the bank at the overnight index-rate, provided that desks do not have a consistent liquidity imbalance between borrowing and lending. When dealing with uncollateralized pricing and even if these hypotheses are not fulfilled, the collateralized framework will always provide a reference or theoretical price on top of which funding and credit value adjustments can be added. 

Section \ref{sec:Perspective} shows how the funding (and collateral) currency is chosen for trading. The proposed decomposition method to price cross currency swaps and foreign exchange forwards is presented in sections \ref{sec:PriceFwdStCCS} and \ref{sec:CustomCCS}. Section \ref{sec:Hedging} compares the proposed method with a benchmark using a worked example. Finally, section \ref{sec:Conclusions} concludes.

\section{Choosing valuation perspective}
\label{sec:Perspective}

The three basic structures which will be considered in this paper are the floating rate note (FRN), the resettable or marked-to-market cross currency swap (CCS) and the non-resettable cross currency swap (NCS). They will be respectively denoted by $FRN^{s^C}_{t_0,t_N}$, $CCS^{s^{C_d},s^{C_f}}_{t_0,t_N}$ and $NCS^{s^{C_d},s^{C_f}}_{t_0,t_N}$, where $C_d$, $C_f$ and $C$ are domestic, foreign or any given currency in which legs are denominated, $t_0$ and $t_N$ are the starting and ending dates of the structures, $s^C$ are the spreads added to the floating leg denominated in currency $C$ and $X_i$ is the foreign exchange rate fixing at $t_i$ to convert 1 unit of domestic currency to foreign. According to this notation, a long position on FRN receives the notional at $t_0$ and pays it back at $t_N$ and in between a floating leg of a given tenor frequency plus the spread $s_C$ is paid. A long position on NCS exchanges notionals at $t_0$ (receives 1 domestic and pays $X_0$ foreign), does the reverse notional exchange at $t_N$ and in between, floating payments plus spreads are exchanged (foreign are received and domestic paid). CCS follows the same convention but resets notionals on each payment date\footnote{On top of the NCS floating exchanges, on each payment date, $t_i$, a reverse notional exchange from previous date is performed (pay 1 domestic and receive $X_{i-1}$ foreign) along with an initial notional exchange on current date (receive 1 domestic and pay $X_i$ foreign). Domestic notional exchanges cancel each other and foreign compensate each other and only the difference is exchanged (receive $X_{i-1}-X_i$ foreign).}.

\begin{equation}
NCS_{t_0 ,t_N }^{s^{C_d} ,s^{C_f} }  = FRN_{t_0 ,t_N }^{s^{C_d} }  - X_{t_0} \cdot FRN_{t_0 ,t_N }^{s^{C_f}  } 
  \label{eq:NCS}
\end{equation}

\begin{equation}
NCS_{t_0 ,t_N }^{s^{C_d } ,s^{C_f } }  = CCS_{t_0 ,t_N }^{s^{C_d } ,s^{C_f } }  + \sum\limits_{i = 1}^{N - 1} {\left( {X_{t_i}  - X_{t_{i - 1}} } \right)FRN_{t_i ,t_N }^{s^{C_f } } } 
  \label{eq:NCS_CCS}
\end{equation}

Equation (\ref{eq:NCS}) presents the decomposition of a non-resettable NCS into a sum of two floating rate notes and equation (\ref{eq:NCS_CCS}) shows how a NCS can be decomposed into a resettable CCS plus a sum of floating rate notes, where $X_{t_i}$ denotes the foreign exchange fixing on $t_i$.


Consider the operator, ${\bf V}_{t}^{C_d} [ {{\bf 1}_{\left\{ {t = T } \right\}}^{C_f } } ]$, where the indicator function, ${\bf 1}_{\left\{ t=T \right\}}^{C_f}$, represents a cash flow payment in foreign currency $C_f$ at time $T$. This operator is defined as the present value (at time $t$) in domestic currency $C_d$, of a cash flow denominated in $C_f$ when it is funded in $C_d$. Equation (\ref{eq:EIndicator2C}) shows how this operator is defined depending on whether the funding and cash flow currencies are the same. $DF_{t,T}^{C}$ is the OIS discount factor between $t$ and $T$ of currency $C$ and $X_{t,T}$ is the forward exchange rate at time $t$ to change one unit of currency $C_d$ to $C_f$ at time $T$ (these forward exchange rates are interpolated according to the end of section \ref{sec:PriceFwdStCCS}). Therefore, the operator ${\bf V}$ is simply defined as OIS discounting when both the cash flow and the funding currencies are the same and when they are different, the operator is defined as the cash flow value converted to the funding currency using the forward foreign exchange and thereafter OIS discounted in that currency. When this operator is applied to a linear combination of cash flows, the result is defined as the linear combination of the operator applied to each cash flow. Cash flows are assumed to be deterministic and no probability measure is defined. In this context, the operator only calculates present values using OIS discounting and it will be assumed that before applying the operator $\bf V$, to FRN, XCS and CCS structures, all contingent index-rate fixings will have been replaced by their forward values so that every cash flow is deterministic before applying the operator.

\begin{equation}
\begin{array}{*{20}c}
   {{\bf V}_t^{C } \left[ {{\bf 1}_{\left\{ {t = T } \right\}}^{C } } \right] = DF_{t,T}^{C } } & {} & {{\bf V}_t^{C_d } \left[ {{\bf 1}_{\left\{ {t = T } \right\}}^{C_f } } \right] = X_{t,T}^{ - 1} DF_{t,T}^{C_d } }  \\
\end{array}
  \label{eq:EIndicator2C}
\end{equation}

Since current regulation enforces inter-bank trades to be liquidated through Clearing Counter Parties (CCP) such as London Clearing House or exchange collateral in form of variation margin for over-the-counter structures and collateral interest payments are indexed to overnight rates such as EONIA, Fed Funds, Sonia, etc, it is well accepted in the market place to use OIS discounting as the reference funding rate in each currency. In the ISDA Master Agreement which defines the netting set in case of default for a given counterparty, the CSA (Credit Support Annex) specifies the currency in which collateral is exchanged for a set of products. In case of default, the deals associated to each CSA net out in the CSA currency and the collateral balance is taken to compensate for the loss. Therefore, the funding currency is specified by the CSA as the set of deals associated with it exchange collateral in that currency and the funding rate is the OIS. Therefore, the heuristic method takes into account the currency of the collateral, by changing each deterministic cash flow from its own currency to the collateral currency using the foreign exchange forward and thereafter by OIS discounting in that currency according to the right side of equation (\ref{eq:EIndicator2C}).

\begin{equation}
{\bf V}_t^{C_d} \left[ FRN_{t_0 ,t_N }^{s^{C_f} } \right] = \frac{DF_{t,t_0}^{C_d}}{X_{t,t_0}} - \sum\limits_{i = 0}^{N - 1} {\left( {L_{t,t_i}^{C_f}  + s^{C_f} } \right)\tau _i^{C_f} \frac{DF_{t,t_{i + 1}}^{C_d}}{X_{t,t_{i+1}}}}  - \frac{DF_{t,t_N}^{C_d}}{X_{t,t_N}}
  \label{eq:FRNf_DOM}
\end{equation}

\begin{equation}
{\bf V}_t^{C_d} \left[ FRN_{t_0 ,t_N }^{s^{C_d} } \right] = DF_{t,t_0}^{C_d} - \sum\limits_{i = 0}^{N - 1} {\left( {L_{t,t_i}^{C_d}  + s^{C_d} } \right)\tau _i^{C_d} DF_{t,t_{i + 1}}^{C_d}}  - DF_{t,t_N}^{C_d}
  \label{eq:FRNd_DOM}
\end{equation}

Equations (\ref{eq:FRNf_DOM}) and (\ref{eq:FRNd_DOM}) show two examples of how the $\bf V$ operator is defined to calculate the present value expressed and funded in domestic units of an FRN denominated in foreign and domestic currencies. $L_{t,t_i}^{C}$ is the forward Libor index rate of currency $C$ at time $t$ of the period from $t_i$ to $t_{i+1}$ and $\tau_i^{C}$ is the year fraction from $t_i$ to $t_{i+1}$ according to the conventions of currency $C$. See that the result of equation (\ref{eq:FRNf_DOM}) will not be equal to ${\bf V}_t^{C_f} \left[ { FRN_{t_0 ,t_N }^{s^{C_f} } } \right] X_t^{-1}$, unless foreign exchange forwards get aligned with overnight cash and carry arguments as the funding currency is different.

The present values provided by the operator $\bf V$ are indeed a heuristic approximation of the actual price. Convexity corrections or the effect of correlation between Libor and foreign exchange rates are ignored. The errors of this approximation will not have much impact, because the floating cash flows which do not net out in the derivative portfolio will be hedged and therefore swapped into fixed deterministic cash flows which will be properly priced according to the $\bf V$ operator.

\begin{table}[htbp]
    \centering
    \begin{tabular}{|p{6.4cm}||p{6.4cm}|}
    \hline
    \textbf{Four-curve method} & \textbf{Heuristic method} \\
    \hline
    \hline
    \multicolumn{2}{|p{12.8cm}|} {Build FF and EO OIS curves forcing overnight index swaps (OIS) = 0.} \\
    \hline
    \multicolumn{2}{|p{12.8cm}|} {Build U3M \& E6M estimation curves forcing most liquid IRS = 0 and discounting with FF and EO.}  \\
    \hline
    \multicolumn{2}{|p{12.8cm}|} {Build USLIBOR \& EURIBOR estimation curves for rest of tenors forcing tenor swaps = 0.}  \\
    \hline
    Build USD collateralized EUR BASIS discount curve forcing CCS = 0. Calculate FX forward curve with FF and EUR BASIS curves. & Build FX forward curve from mkt (short term) and forcing USD funded heuristic value of CCS = 0 (medium and long term). \\
    \hline
    Build EUR collateralized USD BASIS discount curve from unchanged FX forwards and EO OIS curve. & Build EUR col. CCB curve forcing EUR funded heuristic value of CCS = 0 with unchanged FX forwards. \\
    \hline
    \textbf{EUR col}: use estimation curves, discount with EO and USD BASIS. \newline \textbf{US col}: use estimation curves, discount with FF and EUR BASIS. & Apply heuristic decomposition and valuation funded in collateral currency using derived CCB curve (EUR col) or mkt one (USD col). \\
    \hline
    \textbf{EUR col}: estim. EUR/USD FX fwd with EO \& USD BASIS. \newline \textbf{US col}: estim. EUR/USD FX fwd with FF \& EUR BASIS. & Estimate EUR/USD FX fwd by heuristic interpolation in FX fwd curve (see end of section \ref{sec:PriceFwdStCCS}). \\
    \hline
    \end{tabular}
  \caption{Comparison of four-curve (benchmark) and heuristic approaches.}
  \label{tab:MethodComparison}
\end{table}

Table \ref{tab:MethodComparison} compares the steps for valuation using the heuristic method and what will be considered the benchmark: the four-curve approach. It is described for the particular case of EUR and USD currencies but can be applied to any pair of currencies without loss of generality (for these pair of currencies, the market currency swap is collateralized in USD). See that the first three steps are common and they are all carried out through a bootstrapping process. The first one constructs the OIS curves (``FF'', Fed Funds for USD and ``EO'', EONIA, for EUR) by forcing the overnight index swaps to be equal to zero. The second obtains the estimation curves associated with the most liquid tenors (EURIBOR6M, ``E6M'' in EUR and USLIBOR3M, ``U3M'' in USD) by forcing fixed versus floating interest rate swaps to be equal to zero and the third builds the rest of tenors of the estimation curves by forcing tenor swaps (floating versus floating such as EURIBOR3M versus EURIBOR6M) to be equal to zero (see \cite{Fujii2010} for more details).

The fourth step of the four-curve approach replaces the discount curve of the currency different from the collateral currency of the market CCS by a basis discount curve so that CCS are equal to zero (using the estimation and the OIS discount curve for the funding or collateral currency). Foreign exchange forwards are then calculated from the basis and OIS discount curves. In the heuristic approach, the foreign exchange forward curve is build out of swap point market quotes for the short term and the rest of the curve is estimated by a bootstrapping method forcing to zero the heuristic valuation of the cash flows associated with market cross currency swaps funded in its collateral currency according to equations (\ref{eq:FRNf_DOM}) and (\ref{eq:FRNd_DOM}). The interpolation of FX forwards in between two consecutive maturities is carried out according to the end of section \ref{sec:PriceFwdStCCS}. For the purpose of valuation of the decomposition in section \ref{sec:CustomCCS}, this foreign exchange forward curve does not need to be very exact as the cash flows involved are small.

The fifth step calculates the basis discount curve of the four-curve approach when collateral is not exchanged in the currency of the market CCS. This basis curve is calculated assuming that foreign exchange forwards are constant irrespective of the currency in which collateral is exchanged. As they are calculated from the basis and OIS discount curves, the basis curve can be derived from the foreign exchange forward and the OIS curve of the currency in which collateral is exchanged. For the heuristic approach, the cross currency basis spread (CCB) curve quoted by the market assumes collateralization in the currency defined in CCS contracts (USD for CCS exchanging EUR and USD cash flows). However, if collateralization is carried out in the other currency, it is necessary to derive an alternative EUR collateralized cross currency basis (CCB) curve which will be used in the decomposition of section \ref{sec:PriceFwdStCCS}. This curve is derived by finding the cross currency basis spread to force the heuristic valuation to be equal to zero of CCS for all maturities funded in the other currency from the market convention according to equations (\ref{eq:FRNf_DOM}) and (\ref{eq:FRNd_DOM}). The difference between cross currency basis spreads assuming collateralization in either the currency of the CCS or the other one is usually or the order of a few basis points. 

The sixth step shows the valuation with the four-curve approach depending on the funding currency. Estimation curves are used irrespective of the funding currency, the OIS discount curve is used for the currency in which the operation is funded (the collateral currency) and the basis discount curve is used for the other currency. The heuristic method applies the decomposition of section \ref{sec:CustomCCS} using the market cross currency basis curve if the collateral exchanged is the one specified in CCS contracts or the derived cross currency basis curve if collateral is exchanged in the other currency.

Finally, the last step shows the estimation of the foreign exchange forwards. For the four-curve approach, they are calculated using the OIS and basis discount curves used for valuation (see previous steps) and the heuristic method just interpolates them in the previously calculated foreign exchange forward curve.

\begin{equation}
DF_{t,t_i }^{USBASIS}  = X_{t,t_i }^{ - 1} DF_{t,t_i }^{EO} X_t  \Rightarrow X_{t,t_i }  = X_t \frac{{DF_{t,t_i }^{EO} }}{{DF_{t,t_i }^{USBASIS} }}
  \label{eq:FXFwdUS}
\end{equation}

\begin{equation}
DF_{t,t_i }^{EURBASIS}  = X_{t,t_i } DF_{t,t_i }^{FF} X_t^{ - 1}  \Rightarrow X_{t,t_i }  = X_t \frac{{DF_{t,t_i }^{EURBASIS} }}{{DF_{t,t_i }^{FF} }}
  \label{eq:FXFwdEUR}
\end{equation}

Equations (\ref{eq:FXFwdUS}) and (\ref{eq:FXFwdEUR}) show the equivalence of a discount factor of a basis curve and the valuation of the corresponding cash flow with the heuristic approach. The left hand side of equation (\ref{eq:FXFwdUS}) represents the present value of a USD cash flow funded in EUR (that is why it is discounted with the USD basis curve). The right side shows the USD cash flow converted to EUR through the FX forward and discounted with the EONIA OIS curve to fund it in EUR and converted again back to USD with the foreign exchange spot rate. A similar argument follows from equation (\ref{eq:FXFwdEUR}) when the funding currency is USD. See that foreign exchange forwards must be the same from both equations and both involve the two discount curves used for valuation depending on the funding (collateral) currency. When market foreign exchange forwards match those obtained applying overnight cash and carry arguments (discount factors calculated with OIS curves), the asymmetry of funding advantage disappears and basis curves collapse into OIS. In this situation the heuristic method converges into OIS discounting of each cash flow in the currency in which it is denominated.

In this context, the four-curve and heuristic methods are rather equivalent to each other. Although the four-curve method provides a more general framework for exotic product valuation, the end of section \ref{sec:CustomCCS} shows that for the most frequently traded products (swaps and currency swaps) the heuristic valuation provides more funding flexibility for uncollateralied operations (e.g. trades with non-financial institutions and project financing).

\section{Spread calculation of a zero-value forward start CCS}
\label{sec:PriceFwdStCCS}

This section presents how to calculate the market spread of a forward starting resettable currency swap which makes its present value equal to zero. The calculation is carried out in two steps. In the first step, spreads of forward start CCS whose payment dates belong to the current market schedule (dates on exact multiples of the considered frequency tenor) are calculated through a bootstrapping method. The second step estimates the spreads of customized forward start CCS whose product payment dates are in between that market schedule.

Consider $t_i^{mkt}$ the market schedule of payments considered at present time, $t=t_0^{mkt}$, for a given tenor frequency (e.g. 3 months, 6 months, etc) with year fractions, $\tau_i^{mkt}$, corresponding to periods from $t_i^{mkt}$ to $t_{i+1}^{mkt}$. The function $p(T)$ returns the period number ending at or beyond $T$ within the market shedule ($t_{p(t_i^{mkt})}^{mkt} = t_i^{mkt}$).

\begin{equation}
\begin{array}{l}
 \sum\limits_{i = 1}^{p\left( {t_N } \right)} {s_N^{mkt} \tau _{i-1}^{mkt} DF_{t,t_i }^{C_d } }  =  \\ 
 \sum\limits_{i = 1}^{p\left( {t_M } \right)} {s_M^{mkt} \tau _{i-1}^{mkt} DF_{t,t_i }^{C_d } }  + \sum\limits_{i = p\left( {t_M } \right) + 1}^{p\left( {t_N } \right)} {s_{MN}^{mkt} \tau _{i-1}^{mkt} DF_{t,t_i }^{C_d } }  \\ 
 \end{array}
  \label{eq:SUVmkt_Cd}
\end{equation}

The first step calculates the spread, $s_{MN}^{mkt}$, of forward starting CCS within market schedule applying the bootrapping equation (\ref{eq:SUVmkt_Cd}). This equation shows the relation among the spreads of market spot and forward starting CCS. A quoted spot starting CCS expiring at $t_N$, $CCS_{t,t_N }^{s^{C_d }  = s_N^{mkt} ,s^{C_f }  = 0}$, can be expressed as the composition of a spot starting CCS expiring at $t_M$, $CCS_{t,t_M }^{s^{C_d }  = s_M^{mkt} ,s^{C_f }  = 0}$, and a forward starting CCS, $CCS_{t_M,t_N }^{s^{C_d }  = s_{MN}^{mkt} ,s^{C_f }  = 0}$, starting at $t_M$ and expiring at $t_N$. Equation (\ref{eq:SUVmkt_Cd}) only considers the spread payments because the rest of cash flows are the same between the spot CCS expirying at $t_N$ and the composition of spot and forward start CCS. This equation assumes that the fixed spreads are set on the domestic currency leg (this case corresponds to a domestic currency different from USD). The solution of the first step is the calculation of the unobserved spread, $s_{MN}^{mkt}$, from $s_N^{mkt}$ and $s_M^{mkt}$ which are quoted in the market\footnote{It is assumed that the frequency of the floating legs is 3 months. If the floating leg frequency is different from 3 months in either leg, a synthetic market quote for the spot CCS should be obtained introducing the quotes of tenor swaps.}.

Similarly, $t_i^{prd}$, is the schedule of payments of the customized product starting at $t_U$ and ending at $t_V$ and the function $q(T)$ returns the period number within the product schedule ending at or beyond $T$. The nearest times of the market schedule before and after a given time, $t_H$, are respectively denoted by $t_{\underline H }^{mkt}$ and $t_{\overline H }^{mkt}$. 

\begin{equation}
{\bf V}_t^{C_d } \left[ {FRN_{t_U ,t_V }^{s^{C_d }  = s_{UV}^{OIS} } } \right] - X_t^{-1} {\sum\limits_{j = 1}^{q\left( {t_V } \right)} X_{t,t_{j-1}^{prd}}^{OIS} {\bf V}_t^{C_f } \left[ FRN_{t_{j - 1}^{prd} ,t_j^{prd} }^{s^{C_f }  = 0 }  \right] } = 0
  \label{eq:SUV_SCS_Cd}
\end{equation}

\begin{equation}
X_{t,T }^{OIS}  = X_t \frac{{DF_{t,T }^{C_d } }}{{DF_{t,T }^{C_f } }}
  \label{eq:XSCS}
\end{equation}

The second step starts considering the customized forward start CCS (from $t_U$ to $t_V$) and the two contiguous CCS with dates in the market schedule which have the same time duration and start just before and after the start date of the customized CCS (from $t_{\underline U}^{mkt}$ to $t_{\underline V}^{mkt}$ and from $t_{\overline U}^{mkt}$ to $t_{\overline V}^{mkt}$). For these three structures, the spreads forcing their zero value are calculated assuming OIS discounting of every cash flow in its own currency. For the customized CCS, this implies that $s_{UV}^{OIS}$ is derived to satisfy equation (\ref{eq:SUV_SCS_Cd})\footnote{See that in equation (\ref{eq:SUV_SCS_Cd}), the resettable leg of the CCS has been expressed as the sum of forward starting one period FRN structures.}. Similar equations can be written for the contiguous CCS (replacing $t_U$ by $t_{\underline U}$, $t_V$ by $t_{\underline V}$, $t_j^{prd}$ by $t_j^{mkt}$ and $s_{UV}^{OIS}$ by $s_{\underline U \underline V}^{OIS}$). The foreign exchange forward, $X_{t,T}^{OIS}$, is approximated by equation (\ref{eq:XSCS}) assuming overnight cash and carry arguments based on OIS curves. The solution would give the spreads $s_{UV}^{OIS}$, $s_{\underline U \underline V}^{OIS}$ and $s_{\overline U \overline V}^{OIS}$, which do not include the cross currency basis spread yet.

\begin{equation}
\begin{array}{*{20}c}
   {e_{\underline U \underline V }  = s_{\underline U \underline V }^{mkt}  - s_{\underline U \underline V }^{OIS} } & & {e_{\overline U \overline V }  = s_{\overline U \overline V }^{mkt}  - s_{\overline U \overline V }^{OIS} }  \\
\end{array}
  \label{eq:eUVupdown}
\end{equation}

\begin{equation}
e_{UV}  = \frac{{t_{\overline U }^{mkt}  - t_U }}{{t_{\overline U }^{mkt}  - t_{\underline U }^{mkt} }}e_{\underline U \underline V }  + \frac{{t_U  - t_{\underline U }^{mkt} }}{{t_{\overline U }^{mkt}  - t_{\underline U }^{mkt} }}e_{\overline U \overline V } 
  \label{eq:eUV}
\end{equation}

The second step follows from equation (\ref{eq:eUVupdown}) which calculates the error of the approximation of the spreads of the two contiguous CCS, $s_{\underline U \underline V}^{OIS}$ and $s_{\overline U \overline V}^{OIS}$, obtained from equation (\ref{eq:SUV_SCS_Cd}). Finally, the error made by the heuristic equation (\ref{eq:SUV_SCS_Cd}) for the customized forward starting CCS spread, $s_{UV}^{OIS}$, is interpolated between the errors of the two contiguous CCS according to equation (\ref{eq:eUV}). This interpolation has good accuracy as it involves the pricing error between two well-calculated market points. Therefore, the accuracy is not significantly lost by the error interpolation because it is indeed small. The spread, $s_{UV}^{OIS}$, would be correctly calculated and no correction would be needed when the FX forward is obtained from overnight cash and carry arguments.

\begin{equation}
s_{UV}^{mkt}  = s_{UV}^{OIS}  + e_{UV} 
  \label{eq:sUVmkt}
\end{equation}

Equation (\ref{eq:sUVmkt}) finally obtains the customized forward CCS spread by adding to the approximated spread, $s_{UV}^{OIS}$, the error, $e_{UV}$, which is obtained through linear interpolation between two accurately-estimated errors.

The interpolation of forward foreign exchange rates on a particular date is carried out similarly. They are priced according to the usual cash and carry argument applied to OIS curves according to equation (\ref{eq:XSCS}). The errors between the actual market quotes and this approximation are calculated as in equation (\ref{eq:eUVupdown}) for the foreign exchange quoted dates previous and following the date to interpolate. The error on the interpolation date is approximated with equation (\ref{eq:eUV}). Finally, the interpolated error is added to the $X^{OIS}$ estimation of the foreign exchange forward as in equation (\ref{eq:sUVmkt}).


\section{Pricing customized currency swaps}
\label{sec:CustomCCS}

This section prices a customized resettable currency swap through a decomposition procedure in which the currency swap is expressed as the sum of a forward starting market currency swap (whose zero value spread is calculated according to section \ref{sec:PriceFwdStCCS}) plus some additional small contributions priced according to section \ref{sec:Perspective}. This will allow flexibility to choose the currency from which funding and hedging are carried out. Pricing inaccuracies in these small contributions do not usually have a significant impact in pricing  (it is like a fine tuning) as the big contributions have been expressed in terms of market quotes.

Consider a customized CCS starting at $t_U$ and ending at $t_V$ with customized fixed spreads in both legs $s_{UV}^{C_d}$ and $s_{UV}^{C_f}$. The swap is valued at time $t=t_L$, in between the two product fixing dates $t_{\underline L}^{prd}$ and $t_{\overline L}^{prd}$ where $t_{\underline L}^{prd} < t < t_{\overline L}^{prd}$ ($t_{q(t_{\underline L})} = t_{\underline L}$ and $t_{q(t_{\overline L})}=t_{\overline L}$).

\begin{equation}
\begin{array}{l}
 CCS_{t_U ,t_V }^{s^{C_d }  = s_{UV}^{C_d } ,s^{C_f }  = s_{UV}^{C_f } }  = CCS_{t_{\overline L}^{prd} ,t_V }^{s^{C_d }  = s_{\overline LV}^{mkt} ,s^{C_f }  = 0}  - X_{t_{\underline L}^{prd} } FRN_{t_{\underline L}^{prd} ,t_{\overline L}^{prd} }^{s^{C_f }  = s_{UV}^{C_f } }  \\ 
  + FRN_{t_{\underline L}^{prd} ,t_{\overline L}^{prd} }^{s^{C_d }  = s_{UV}^{C_d } }  + \sum\limits_{j = q\left( {t_{\overline L}^{prd} } \right) + 1}^{q\left( {t_V } \right)} {\left[ {\left( {s_{UV}^{C_d }  - s_{\overline LV}^{mkt} } \right){\bf 1}_{\left\{ {t = t_j^{prd} } \right\}}^{C_d }  + s_{UV}^{C_f } {\bf 1}_{\left\{ {t = t_j^{prd} } \right\}}^{C_f } } \right]}  \\ 
 \end{array}
  \label{eq:CCSdecom}
\end{equation}

Equation (\ref{eq:CCSdecom}) shows the currency swap decomposition. The first term of the right hand side is a forward starting resettable currency swap whose spread, $s_{\overline L V}^{mkt}$, is obtained according to equation (\ref{eq:sUVmkt}) using the method of section \ref{sec:PriceFwdStCCS}. The second and third terms involving FRN structures represent the swap structure of the already started hub period (exchange of notionals and floating payments at the end of the period) and the sum includes the fixed spread, $s_{UV}^{C_f}$, of the foreign curve and the fixed payments of the domestic leg, equal to the difference between the customized and market spreads, $s_{UV}^{C_d}$ and $s_{\overline L V}^{mkt}$.
No valuation has been performed yet, only a payoff decomposition. To do the pricing, three terms are considered: the CCS, the two FRN structures and the last sum of cash flows. The only multi-currency piece of the decomposition is the forward starting CCS whose valuation is zero ($s_{\overline L V}^{mkt}$ is calculated to satisfy this condition). The rest of the pieces involve single currency fixed cash flows. The two FRN cash flows $CF^{FRN}$ in equation (\ref{eq:FRNvsFRN})\footnote{The FRN cash flows in equation (\ref{eq:FRNvsFRN}) change signs with respect to equation (\ref{eq:CCSdecom}) because the notation for a long position of FRN was defined returning the notional and paying the floating rate at expiry.}, must be priced jointly applying either the operator ${\bf V}_t^{C_d}[\cdot]$ or ${\bf V}_t^{C_f}[\cdot]$ assuming a common funding currency, either domestic or foreign (see that equation (\ref{eq:FRNvsFRN}) has replaced contingent index fixings by their forward values before applying the operator). This common funding avoids the inconsistency between the market forward foreign exchange rate, $X_{t,t_{\overline L}^{prd}}$ and the estimated foreign exchange rate, $X^{OIS}_{t,t_{\overline L}^{prd}}$, of equation (\ref{eq:XSCS}) using OIS discount factors\footnote{This inconsistency could allow for an arbitrage, as these cash flows are big enough (they involve a Libor rate instead of a spread) and the term between their fixing and payment is very liquid (just 3 months).}. Equation (\ref{eq:EFRNvsFRN}) shows the joint valuation of these two FRN cash flows which must be funded in the same currency (usually the collateral currency). Foreign exchange forwards are interpolated according to the end of section \ref{sec:PriceFwdStCCS}.

\begin{equation}
\begin{array}{l}
 {CF}^{FRN}  = CF^{C_f } {\bf 1}_{\{ {t = t_{\bar L}^{prd} } \}}^{C_f }  - CF^{C_d } {\bf 1}_{ \{ {t = t_{\bar L}^{prd} } \}}^{C_d } \\
 CF^{C_d} = [ {1 + ( {L_{t,t_{\underline L}^{prd} }^{C_d }  + s_{UV}^{C_d } })\tau _{q( {t_{\underline L}^{prd} })}^{prd} }]  \\
 CF^{C_f} = X_{t,t_{\underline L}^{prd} } [ {1 + ( {L_{t,t_{\underline L}^{prd} }^{C_f }  + s_{UV}^{C_f } })\tau _{q( { t_{\underline L}^{prd} })}^{prd} }] \\
\end{array}
  \label{eq:FRNvsFRN}
\end{equation}

\begin{equation}
\begin{array}{l}
 {\bf V}_t^{C_d } [CF^{FRN} ] = (CF^{C_f } X_{t,t_{\bar L}^{prd} }^{ - 1}  - CF^{C_d } )DF_{t,t_{\bar L}^{prd} }^{C_d }  \\ 
 {\bf V}_t^{C_f } [CF^{FRN} ] = (CF^{C_f }  - CF^{C_d } X_{t,t_{\bar L}^{prd} } )DF_{t,t_{\bar L}^{prd} }^{C_f }  \\ 
 \end{array}
  \label{eq:EFRNvsFRN}
\end{equation}

Similarly, the last spread sum of equation (\ref{eq:CCSdecom}) should be jointly funded in the collateral currency and for uncollateralized operations, the currency in which it is decided to fund them. Independently of the funding currency chosen for valuation, every price must be converted to the common currency in which pricing is carried out with the foreign exchange spot. This heuristic method allows choosing the funding currency of these additional cash flows which do not belong to the cross currency swap.

%

Non-resettable currency swaps can be priced according to the decomposition of equation (\ref{eq:NCS_CCS}). The first term on the right hand side is a customized resettable cross currency swap which can already be priced. The second term is a sum of floating rate notes which can be priced applying operator $\bf V$ assuming funding in the currency of the collateral or if the operation is not collateralized, the cheapest or most convenient currency to fund it. These discounted cash flows must be converted to the currency in which pricing is accomplished using the corresponding foreign exchange spot.

As it has already been mentioned, if collateralized operations are considered, a common funding must be chosen equal to the collateral currency to allow for valuation agreement with the counterparty with which collateral is exchanged. However, if an uncollateralized transaction is considered, the decomposition of equation (\ref{eq:CCSdecom}) allows for more flexibility to fund the remaining pieces apart from the currency swap depending on trading convenience and preferences. A typical example happens when a European bank hedges cross currency exposure between USD and EUR with other European institutions under a CSA agreement in EUR. Market currency swaps are collateralized in USD but hedges are closed with institutions which quote them as collateralized in EUR. In this situation, the four curve method requires everything to be collateralized in EUR. However, the heuristic method will collateralize the currency swap in EUR but the remaining pieces of the decomposition may be funded in the cheapest or most convenient currency. If funding in EUR is cheaper than in USD (e.g. negative cross currency spreads), USD cash flows will be cheaper to fund in EUR if they are payed and in USD if received. In addition, if a particular institution has better access and conditions to enter into OIS swaps collateralized in USD rather than EUR (e.g. cleared OIS swaps may only be available with USD collateral), the four curve method would not allow to use them as hedges. Therefore the heuristic method allows a better customization of the valuation with the actual hedge products. This is the main contribution of this method. 

\section{Worked example}
\label{sec:Hedging}

This section presents a worked example comparing the two methods with the potential advantage of the heuristic method. The example illustrates the case in which a European institution hedges currency exposure between EUR and USD with other banks under a CSA agreement that exchanges collateral in EUR. The four-curve method funds everything in domestic currency (EUR) and the heuristic method funds the currency exposure in EUR but the rest in USD. The comparison will be done for a resettable (CCS) and a non-resetable (NCS) currency swap, whose legs are denominated in USD and EUR, spreads added to the EUR floating leg of -5.75 basis points and notional amount of 100 million EUR. Market data has been taken on January 29th 2014, with spot and forward foreign exchange rates in USD per EUR of \{Spot: 1.3533, 1y: 1.3543, 2y: 1.3610, 3y: 1.3741, 4y: 1.3928, 5y: 1.4143, 7y: 1.4589, 10y: 1.5145\}, and a cross currency basis spread curve in basis points added to the EUR floating leg and collateralized in EUR of \{1y: -4, 2y: -5.5, 3y: -6.25, 4y: -7, 5y: -7, 7y: -6.75, 10y: -5.75, 15y: -4.75, 20y: -4.5\} with zero spread on the USD floating leg.

The cross currency quotes are provided by an institution with a CSA in EUR\footnote{The sensitivity propagation method used by the system in which the four-curve method is implemented can only recalibrate the set of curves chosen for valuation. If valuation considers EUR collateral, then market CCS are valued with those set of curves when recalibration is carried out and therefore sensitivities are provided as if collateralized in EUR.}. Five curves are considered: EONIA (``EO''), 3 month Euribor (``E3M''), Federal Funds (``FF''), 3 month US Libor (``U3M'') and cross currency basis spread (``CCB'').

\begin{table}[htbp]
    \centering
    \begin{tabular}{r|rrrrr||rrrrr|}
    CCS   & \multicolumn{ 5}{c||}{Four-curve method} & \multicolumn{ 5}{c|}{Heuristic method}  \\
    \hline
    {\bf Mat} & EO & E3M & FF    & U3M & CCB & EO & E3M & FF    & U3M & CCB \\
    \hline
    {\bf 1y}  & 0     & 0     & 0     & 0     & 0     & 0     & 0     & 0     & 0     & 0 \\
    {\bf 5y}  & 0     & 0     & 0     & 0     & -1    & 0     & 0     & 0     & 0     & 0 \\
    {\bf 9y}  & 0     & 0     & 0     & 0     & -2    & 0     & 0     & 0     & 0     & 0 \\
    {\bf 10y} & 0     & 0     & 0     & 0     & 108   & 0     & 0     & 0     & 0     & 96 \\
    \hline
    {\bf NPV} &  &      &       & {\mbox{\euro}} & 0  &  &       &       & {\mbox{\euro}} & 0 \\
    \hline
    \end{tabular}
  \caption{10 year CCS NPV and deltas (thousand EUR) for four-curve and heuristic methods moving each curve by one basis point.}
  \label{tab:CCSdelta}
\end{table}

\begin{table}[htbp]
    \centering
    \begin{tabular}{r|rrrrr||rrrrr|}
    NCS   & \multicolumn{ 5}{c||}{Four-curve method} & \multicolumn{ 5}{c|}{Heuristic method}  \\
    \hline
    {\bf Mat} & EO & E3M & FF    & U3M & CCB & EO & E3M & FF    & U3M & CCB \\
    \hline
    {\bf 1y} & 0     & 0     & 0     & 0     & 0     & 0     & 0     & 0     & 0     & 0 \\
    {\bf 5y} & -1     & 1    & 0     & 0     & 0     & 0     & 0     & -1    & 1     & 0 \\
    {\bf 9y} & -2     & 2    & 0     & 0     & 0     & 0     & 0     & -2    & 2     & 0 \\
    {\bf 10y} & 15   & -15    & 0     & 0     & 97    & 0     & 0     & 14    & -14   & 96 \\
    \hline
    {\bf NPV} &  &      &       & {\mbox{\euro}} & -58      &  &       &       & {\mbox{\euro}} & -120.34   \\
    \hline
    \end{tabular}
  \caption{10 year NCS NPV and deltas (thousand EUR) for four-curve and heuristic methods moving each curve by one basis point.}
  \label{tab:NCSdelta}
\end{table}

Tables \ref{tab:CCSdelta} and \ref{tab:NCSdelta} show NPV (net present value) and delta sensitivities in thousand EUR of each curve for 10 year CCS and NCS under rate movements of one basis point of the four-curve and heuristic methods. The CCS has zero NPV (net present value) for both pricing methods as they are calibrated to market. The NPV difference for the NCS between both methods (see table \ref{tab:NCSdelta}) is 62.26 thousand which reflects the pricing difference of using USD instead of EUR funding for the non-cross currency risk. For the position shown in table \ref{tab:NCSdelta} the heuristic method provides a more negative NPV. This means that EUR funding providing the price given by the four-curve method would be cheaper. However, for the opposite position the USD funding would provide a higher NPV and would be preferable.

The four-curve approach calculates sensitivities by finite differences using a propagation algorithm which before pricing any perturbed scenario again, re-calibrates the basis USD discount curve after moving either interest or foreign exchange rates.

For the CCS of table \ref{tab:CCSdelta}, both methods provide zero sensitivities to every curve except for the cross currency curve, ``CCB''. This is as expected because the heuristic algorithm decomposes the CCS in terms of direct quotes of CCB curve and the four-curve uses the propagation method to calculate sensitivities. Some minor sensitivities appear in CCB curve for the four-curve method for 5 and 9 years. There is a slight mismatch for the four-curve method at 10 year maturity (108 versus 96) which does not happen for the NCS of table \ref{tab:NCSdelta}. This mismatch is much lower for a 5 year CCS (52 versus 50.1) and it might be explained by slight numerical problems of the propagation sensitivity algorithm (this is avoided with the proposed method).

According to equation (\ref{eq:NCS_CCS}), the NCS is decomposed in a CCS plus a series of FRN structures denominated in USD. As the forward foreign exchange curve is increasing, the floating cash flows are always paid. Table \ref{tab:NCSdelta} shows the sensitivities of the NCS. The heuristic method funds those cash flows in USD and yields sensitivities to them for ``FF'' and ``U3M''\footnote{Rising ``U3M'' curve increases USD paid cash flows and so the biggest sensitivity is negative as more has to be paid. Rising ``FF'' curve decreases USD payments providing a positive sign for the biggest sensitivity.} curves as they are funded and discounted with ``FF'' curve. See that sensitivities to ``EO'' and ``E3M'' curves are zero as every EUR cash flow is incorporated into the CCS (the sensitivity appears in the CCB curve). The four-curve method only provides sensitivities  to EUR curves with the same sign as the corresponding USD sensitivities of the proposed method\footnote{Rising ``EO'' curve reduces value of EUR leg of the calibrated CCS, which has to be compensated by an increase of USD basis discount curve after propagation (market CCS re-calibration) providing a positive sensitivity (the same sign of the ``FF'' curve for the heuristic). Rising ``E3M'' curve lowers USD basis discount curve after propagation, yielding a negative sensitivity (opposite sign of ``FF'' curve for heuristic).}. The sensitivities are of the same order for both ``EO'' and ``FF'' curves as they have similar interest rate levels.

\begin{equation}
\begin{array}{l}
 \Delta ^\$   = \frac{{\partial \left( {V^\$  X_t^{ - 1} } \right)}}{{\partial \left( {X_t^{ - 1} } \right)}} =  - \Delta ^{\mbox{\euro}} X_t + V^\$   \\ 
 \Delta ^{\mbox{\euro}}  = \frac{{\partial V^\$  }}{{\partial X_t}} =  - \Delta ^\$  X_t^{ - 1}  + V^\$  X_t^{ - 1} \\ 
 \end{array}
  \label{eq:FXdelta}
\end{equation}

Equation (\ref{eq:FXdelta}) shows foreign exchange deltas as seen from Europe ($\Delta^{\$}$ USD) and the US ($\Delta^{\mbox{\euro}}$ EUR), where $V^{\$}$ is the price in USD and $X_t$ is the spot foreign exchange rate (USD per unit of EUR). Pricing systems based in Europe usually report $\Delta^{\mbox{\euro}} - V^{\$} X_t^{-1}$ as FX delta, because the deal premium is already in EUR and this amount has to be subtracted from the total amount of Euros to cancel, $\Delta^{\mbox{\euro}}$, in order to be delta hedged. See that this cancelling EUR transaction, $\Delta^{\mbox{\euro}} - V^{\$} X_t^{-1} = -\Delta^{\$} X_t^{-1}$, is carried out against a quantity of USD, $-\Delta^{\$}$, which cancels the open US sensitivity, $\Delta^{\$}$. For a US investor, the delta reported by the system would be $\Delta^{\$} - V^{\$}$.

\begin{table}[htbp]
    \centering
    \begin{tabular}{r|rr||rr|}
     & \multicolumn{ 2}{|c||}{Four-curve} & \multicolumn{ 2}{c|}{Heuristic} \\
    \hline
    FX Delta                   & CCS   & NCS    & CCS   & NCS \\
    \hline
    $\Delta^{\mbox{\euro}}$ EUR & 39    & -1,380 & -37   & -2,623 \\
    $\Delta^{\$}$           USD & 26    & 1,884  & 52    & 3,417  \\
    \hline
    \end{tabular}
  \caption{10 year foreign exchange deltas (thousands) of CCS and NCS for four-curve and proposed methods.}
  \label{tab:FXdelta}
\end{table}

Table \ref{tab:FXdelta} shows the foreign exchange deltas of CCS and NCS for four-curve and heuristic methods according to equation (\ref{eq:FXdelta}). Values corresponding to $\Delta^{\mbox{\euro}}$ are in thousand EUR and to $\Delta^{\$}$ in thousand USD. To understand the differences with respect to the two products and methods, see that for the heuristic method under the assumption that terms not belonging to the CCS (the foreign exchange delta is not affected by the funding currency of the CCS) are funded in USD, the foreign exchange deltas are given by equation (\ref{eq:FXdeltaUSDfund}), where $\Delta_{CCS, USD}^{\$}$ stands for $\Delta^{\$}$ for the resettable currency swap whose additional cash flows of the decomposition are funded in USD. See that for the CCS, only the first 3 month period contributes\footnote{According to the decomposition of equation (\ref{eq:CCSdecom}) the terms which affect the foreign exchange delta of equation (\ref{eq:FXdelta}) are the FRN denominated in foreign currency (USD), $- X_{t_0} FRN_{t_0 ,t_0+3m }^{s^{\$}=0}$ and the sum of foreign spreads, $s_{UV}^{C_f}$, which are zero in this example.}. For the NCS, the whole 10 year period contributes\footnote{According to the decomposition of equation (\ref{eq:NCS_CCS}) the CCS contributes with the first period, $- X_{t_0} FRN_{t_0 ,t_0+3m }^{s^{\$}=0}$ and from the sum, only the term $-X_{t_0}FRN_{t_1,t_0+10y}^{s^{\$}=0}$ contributes as the foreign exchange rate has already been fixed. These two terms together are equivalent to $-X_{t_0}FRN_{t_0,t_0+10y}^{s^{\$}=0}$.}.

\begin{equation}
\begin{array}{l}
 \Delta _{CCS, USD}^\$   = {\bf V}_t^\$  \left[ { - X_{t_0 } FRN_{t_0 ,t_0  + 3m}^{s^\$   = 0} } \right]  \\ 
 \Delta _{NCS, USD}^\$   = {\bf V}_t^\$  \left[ { - X_{t_0 } FRN_{t_0 ,t_0  + 10y}^{s^\$   = 0} } \right] \\ 
 \end{array}
  \label{eq:FXdeltaUSDfund}
\end{equation}

Foreign exchange deltas of CCS are much smaller than those of NCS because for the CCS only a period of 3 months contributes, whereas for the NCS, the whole 10 year period contributes. CCS EUR deltas ($\Delta^{\mbox{\euro}}$) change signs between four-curve and proposed methods (39 versus -37), because both CCS premium and delta are small and comparable (see equation (\ref{eq:FXdelta}) to relate EUR and USD deltas).

Equation (\ref{eq:FXdeltaEURfund}) shows the foreign exchange deltas when the heuristic method funds the remaining terms apart from the CCS component in EUR. See that the operator ${\bf V}_t^{\mbox{\euro}} [ \cdot ]$ is used to represent that cash flows are converted with foreign exchange forwards to EUR and thereafter discounted with curve ``EO''. If this heuristic valuation is differentiated with respect to $X_{t}^{-1}$, the $X_{t}^{-1}$ factors of the foreign exchange forwards will disappear which is equivalent to multiply deltas by $X_t$ as shown in equation (\ref{eq:FXdeltaEURfund}). If these deltas are calculated with the heuristic method as shown in equation (\ref{eq:FXdeltaEURfund}) in thousand USD, see that they are similar to those calculated by the four-curve method (see second row, left side of table \ref{tab:FXdelta}).

\begin{equation}
\begin{array}{l}
  \Delta^{\$}_{CCS,EUR} = X_t {\bf V}_t^{\mbox{\euro}} [ - X_{t_0} FRN_{t_0,t_0+3m}^{s^{\$}=0}  ]  = 26    \\ 
  \Delta^{\$}_{NCS,EUR} = X_t {\bf V}_t^{\mbox{\euro}} [ - X_{t_0} FRN_{t_0,t_0+10y}^{s^{\$}=0} ] = 1,906  \\ 
 \end{array}
  \label{eq:FXdeltaEURfund}
\end{equation}

See that foreign exchange deltas differ depending on the funding currency chosen by the heuristic method mainly because the valuation of the last notional payment of the FRN may differ depending on whether it is discounted with ``FF'' or converted to EUR and discounted with curve ``EO''. As EUR funding has advantage compared to USD, the EUR funded delta is lower.

\section{Conclusions}
\label{sec:Conclusions}

A heuristic present value concept for multi-currency pricing and hedging has been proposed which naturally allows choosing the funding and therefore the valuation collateral currency. This concept is applied to the valuation of cross currency swaps by decomposing them into a major component with the cross currency basis risk plus minor residual components. This method has been compared with the four-curve method which is the current benchmark. For collateralized operations with funding managed in the collateral currency both methods are rather equivalent. For uncollateralized operations, the heuristic method allows more optionality to choose the funding of the components without cross currency basis risk to achieve either a cheaper funding or more connected one to the hedging products. Although the four-curve method provides a more general framework for exotic valuation, the heuristic method provides more funding flexibility for the most frequently traded products. The heuristic method converges to OIS discounting in each currency when foreign exchange forwards follow overnight cash and carry arguments.

\begin{table}[htbp]
    \centering
    \begin{tabular}{|p{13.3cm}|}
    \hline
    {\bf Acknowledgement}: the authors want to thank E. Cuesta and G. Montesinos for their support, feedback and clarifying discussions. \\
    \hline
    \end{tabular}
\end{table}


\begin{thebibliography}{30}

\bibitem{Bianchetti2009b}
F. Ametrano and M. Bianchetti, ``Bootstrapping the Illiquidity: Multiple Yield Curves Construction for Market Coherent Forward Rates Estimation'', Modeling Interest Rates: Last advances for Derivatives Pricing, Ed. F. Mercurio, Risk Books, 2009.

%
%
%
%
\bibitem{Cuesta2012}
Cuesta E., ``Cross currency basis formula'', Working paper, February 2012, URL: ``http://papers.ssrn.com/sol3/papers.cfm?abstract\_id =2014762''.
%
%
%
%
%
%
%
%
%
%
\bibitem{Fujii2010}
Fujii M., Shimada Y., Takahashi A., ``A note on construction of multiple swap curves with and without collateral'', FSA Research Review, Vol. 6, March 2010.
%
\bibitem{Boenkost2005}
Boenkost W., Schmidt W., ``Cross currency swaps valuation'', Working paper, May 2005, URL: ``http://papers.ssrn.com/ sol3/papers.cfm?abstract\_id=1375540''.
%
%

\end{thebibliography}
\end{document}